# Examining the impact of data quality and completeness of electronic health records on predictions of patients' risks of cardiovascular disease

Yan Li[1], Matthew Sperrin[1], Glen P. Martin[1], Darren M Ashcroft[2,3], Tjeerd Pieter van Staa[1,4,5]

[1]Health e-Research Centre, Farr Institute, School of Health Sciences, Faculty of Biology, Medicine and Health, the University of Manchester, Manchester Academic Health Sciences Centre (MAHSC), Oxford Road, Manchester, M13 9PL, UK
[2]Centre for Pharmacoepidemiology and Drug Safety, School of Health Sciences, Faculty of Biology, Medicine and Health, University of Manchester, Oxford Road, Manchester, M13 9PL, UK
[3]NIHR Greater Manchester Patient Safety Translational Research Centre, School of Health Sciences, Faculty of Biology, Medicine and Health, University of Manchester, Oxford Road, Manchester, M13 9PL, UK
[4]Utrecht Institute for Pharmaceutical Sciences, Utrecht University, Utrecht, Netherlands
[5]Alan Turing Institute, Headquartered at the British Library, London, UK

Corresponding author: Tjeerd van Staa, tjeerd.vanstaa@manchester.ac.uk

**Abstract**

**Objective**

To assess the extent of variation of data quality and completeness of electronic health records and impact on the robustness of risk predictions of incident cardiovascular disease (CVD) using a risk prediction tool that is based on routinely collected data (QRISK3).

**Design:**

Longitudinal cohort study.

**Setting**

392 general practices (including 3.6 million patients) linked to hospital admission data.

**Methods**

Variation in data quality was assessed using Sáez's stability metrics quantifying outlyingness of each practice. Statistical frailty models evaluated whether accuracy of QRISK3 predictions on individual predictions and effects of overall risk factors (linear predictor) varied between practices.

**Results**

There was substantial heterogeneity between practices in CVD incidence unaccounted for by QRISK3. In the lowest quintile of statistical frailty, a QRISK3 predicted risk of 10% for female was in a range between 7.1% and 9.0% when incorporating practice variability into the statistical frailty models; for the highest quintile, this was 10.9%-16.4%. Data quality (using Saez metrics) and completeness were comparable across different levels of statistical frailty. For example, recording of missing information on ethnicity was 55.7%, 62.7%, 57.8%, 64.8% and 62.1% for practices from lowest to highest quintiles of statistical frailty respectively. The effects of risk factors did not vary between practices with little statistical variation of beta coefficients.

**Conclusions**

The considerable unmeasured heterogeneity in CVD incidence between practices was not explained by variations in data quality or effects of risk factors. QRISK3 risk prediction should be supplemented with clinical judgement and evidence of additional risk factors.

# Introduction

Cardiovascular disease (CVD) has been the most common cause of death around the world for decades[1]. The prevention of CVD through targeting treatment to high risk patients is recommended in many international guidelines[1-2]. Risk prediction models are now an important part of CVD prevention strategies[3]. Many CVD risk prediction models have been developed around the world[4], including the Framingham risk score (FRS)[5] in the USA, QRISK3[6] in the UK and ESC HeartScore in Europe[7]. These models were developed by fitting statistical survival models (e.g. Cox model[8]) incorporating CVD risk factors on longitudinal patient cohorts. Specifically, QRISK was first developed in 2008 using routinely collected electronic health records (EHRs) from 355 general practices included in the QResearch database[6]. It considered age, sex and CVD risk factors such as body mass index (BMI) and smoking status. A recent update, QRISK3, incorporated more risk factors, such as variation in systolic blood pressure[6].

A previous study has found that QRISK3 scores that are derived from EHRs can have limited generalisability and accuracy, as they do not account for the substantive heterogeneity between different general practices[9]. Considerable changes in the individual risk estimates occurred when taking into account the heterogeneity between different general practices, and this effect cannot be explained by practice random variability[9]. Additionally, this study found that a CVD risk of 10% over 10 years as predicted by QRISK3 could change by over absolute 13% in a model that also incorporated variability between sites. Heterogeneity between sites may be related to either data quality (mainly including variation of missingness and coding[10]) or unadjusted underlying practice heterogeneities (variation of patient case mix and association between outcome and predictors[11]). However, it is unknown which of these influences contribute to the observed effects of practice variability on individual risk prediction. Therefore, the objective of this study was to assess the extent of variation of data quality and completeness of electronic health records and impact on the robustness of risk predictions of cardiovascular disease (CVD) using QRISK3. The QRISK3 model is recommended to be used in UK general practice and is now also accessible for members of the public[12, 13].

## Methods

The study used data from approximately 3.6 million anonymised patient records derived from 392 general practices from the Clinical Practice Research Datalink (CPRD GOLD), which had been linked to Hospital Episode Statistics (HES), Office for National Statistics (ONS) mortality records and Townsend deprivation scores[6]. CPRD GOLD is a representative demographic sample of the UK population in terms of age, gender and ethnicity[14]. Overall, CPRD includes data on about 6.9% of the UK population. The linkages to other datasets such as HES or ONS provide additional patient information about secondary care, specific disease and cause-specific mortality[14]. CPRD includes patients' electronic health records from general practice capturing detailed information such as demographics, symptoms, tests, diagnoses, prescribed treatments, health-related behaviours and referrals to secondary care[14]. CPRD data has been widely used for public health research[15], including an external validation of the QRISK2 model[16].

The study used the same patient population as described in a previous study[9], and used similar selection criteria and risk factors to QRISK3[6]. The follow-up of patients started at the date of the patient's registration with the practice, 25th birthday, or January 1 1998 (whichever latest), and ended at the date of death or CVD outcome, the date of leaving the practice, end of study window or last date of data collection (whichever earliest). The index date for measurement of CVD risk was randomly chosen from the total period of follow-up[17]. This study used a random index date, as it captures time-relevant practice variability with a better spread of calendar time and age[18]. The use of a random index date was the only difference with the original QRISK3 studies[6]. The main inclusion criteria for the study population were aged between 25 and 84 years, with no CVD history or any statin prescription prior to the index date. Patients were censored at the date of the statin prescription if received during follow-up.

There were four analysis parts in this study. The first measured data quality and completeness in each of the different practices. Second, we evaluated the heterogeneity between practices in CVD incidence that was not taken into account in the development of QRISK3. This analysis addressed the miscalibration of QRISK3 at practice level which can be described as the closeness (accuracy) of the QRISK3 prediction to the observed CVD incidence in each practice. Unmeasured heterogeneity between groups is also known as statistical frailty, which can be modelled in regression analyses[19]. Bootstrap resampling with 1000 times was used to quantify the confidence intervals of individual risk predictions in the random intercept model for patients who had a QRISK3 predicted risk of 10%. The level of unmeasured heterogeneity in CVD incidence (statistical frailty) for each practice was used to stratify practices into quintiles. Third, we evaluated whether the effects of the QRISK3 risk factors (i.e., the overall linear predictor) varied between practices (i.e., whether the beta coefficients varied). This variation in the linear predictor between practices could occur in case of unmeasured effect moderators for CVD incidence or differences in data recording/misclassification of risk factors. Finally, we compared data quality across different levels of statistical frailty.

Several indicators of data quality were used in this study to measure the variation in coding between general practices. First, the percentages of missing records were measured for the variables ethnicity, systolic blood pressure (SBP), body mass index (BMI), cholesterol, high-density lipoprotein (HDL), ratio of cholesterol and HDL, smoking status and Townsend score for deprivation. Second, two metrics as proposed by Sáez[20] were used to measure the multidimensional variability (stability) in data quality across practices. The proposed metrics quantified the variability in the probability distribution functions of practices. Variation of coding was measured by the distribution-dissimilarity (quantified by Sáez's metrics) of CVD risk factors and their missingness among practices. Inconsistent coding of clinical data could result in misclassification and different distributions of the variable among practices, which may influence risk prediction. This effect of misclassification was considered using Saez metric to measure the distribution-dissimilarity of all coded clinical variables including atrial fibrillation, chronic kidney disease, erectile dysfunction, angina or heart attack in a 1st degree relative < 60, migraines, rheumatoid arthritis, systemic lupus erythematosus, severe mental illness, type 1 diabetes and type 2 diabetes. Sáez's metric[20], which was based on Jensen–Shannon divergence[21] measured the distribution-dissimilarity of variables across practices. Specifically, source probabilistic outlyingness (SPO) can be thought of as a measure of how different a practice is from the average practice in terms of distribution of variables. SPO ranges from 0 to 1 measuring the extent of outlyingness of the variables' distribution. A variable with a SPO close to 1 means that the distribution of the variable in the practice is more different from the overall average indicating the outlyingness of coding. Further technical details about the Sáez's metric are provided in the eAppendix 2.

The unmeasured heterogeneity between practices in CVD incidence was evaluated by fitting a Cox proportional hazards model that included a statistical frailty term on its intercept (this type of model is also known a random intercept Cox model). The outcome of interest was the time to CVD onset. The linear predictor of QRISK3 (sum of the multiplication of beta coefficients and predictors) was used as an offset (i.e. coefficient fixed at one) to calculate the statistical frailty for each practice[19].

The variation between practices in the effects of the QRISK3 risk factors was evaluated by also adding a single frailty term to the beta coefficients of the QRISK3 linear predictor (known as a mixed effects Cox model[22]). This model calculated a random slope for the QRISK3 risk factors in each practice (assuming fixed effects and independent random effects of the QRISK3 linear predictor) in addition to the random intercept (assuming unmeasured heterogeneity in CVD incidence between practices). The random slopes and intercepts were calculated separately for each gender as QRISK3 has separate model formula for each gender[23].

The effects of practices' random slopes on individual risk prediction were visualised by estimating the difference of individual CVD risk predictions in the random slope model to that of the random intercept model. The range of individual risk predictions were calculated from the random slope model. Using a QRISK3 risk of 10%, a random slope and a random intercept were randomly drawn from a Gaussian distribution based on the variation of the random slope and random intercept calculated from this study's original cohort and the predicted risk was estimated (this was repeated one million times). The difference of the predicted CVD risk when the same patient was from practices with the same random intercept, but different random slope was visualised. Two hypothetical variations in random slopes (0.03 and 0.1) were used as reference lines. The variation in random slopes of 0 indicates that there was no variation in the effects of CVD risk factors between practices. The variation of 0.03 was chosen as reference because a previous study found that this variation in the random effects of the intercept[9] resulted in large differences in individual risk predictions (a QRISK3 predicted risk of 10% would change in the random effects model to a range from 7.2% to 13.7%).

Finally, practices were grouped by quintiles of statistical frailty and data quality metrics were estimated for each quality indicator. The random intercept Cox models estimated the level of statistical frailty for a QRISK3 predicted CVD risk of 10% (over 10 years). The mean and standard deviation of each CVD risk factors were summarised. Sáez's metric for the CVD risk factors and their missingness were plotted against the percentile of practice frailty to show possible correlations and the Pearson correlation coefficients were calculated[24]. Practice statistical frailty was also plotted against the percentile of the mean (for continuous variables) or percentage (for categorical variable) of CVD risk factors at practice level and their corresponding Sáez's metric using a Beeswarm plot[25]. This aims to identify any correlation between them and practice statistical frailty was plotted with value 1 as a reference line (red line). Beeswarm plots visualise the distribution by plotting practices as separate dots in each bin, so it has benefit to highlight individual points in distribution comparing to classical distribution-visualisation such as histograms.

The statistical software R version 3.4.2[26] with package "coxme"[27] was used to model the data; SAS 9.4 was used in data preparation, missing value imputation and visualisation. Multiple imputation using Markov chain Monte Carlo (MCMC) method with monotone style[28] was used to impute missing values before model fitting. Ten imputed datasets were created with pooling of the results based on the averages.

**Results**

There were 3,630,818 patients included in the study cohort, 103,350 of which had a CVD event in the 10 years after the index date. Overall, for patients with QRISK3 predicted risk of 10%, the 95% range of predicted risks were between 7.2% (95%CI: 7.15%~7.34%) and 13.7% (95%CI: 13.5%~13.9%) with the random intercept model (which incorporated practice heterogeneity into the model). Table 1 shows the differences between the predictions by the QRISK3 and random intercept models (statistical frailty) for patients with a QRISK3 prediction of 10% with practices classified into quintiles of practice statistical frailty. Practices in the lowest quintile had predicted CVD risks at 10 years between 7.1% and 9.0% in the random intercept model for females compared to a predicted risk of 10% with QRISK3. For males, this was 6.1% and 9.0%. For practices in the highest quintile, QRISK3 predictions underestimated CVD risks compared to the random intercept model with predicted risks between 10.9% and 16.4% (for males, this was 10.9% and 15.5%). As shown in Table 1, a practice statistical frailty below 1 indicated that QRISK3 overestimated CVD risk and above 1 underestimated CVD risk compared with the random intercept models.

Table 2 compares the baseline characteristics between practices with different levels of statistical frailty (i.e., mean difference between individual risk predictions by QRISK3 and random intercept models). For example, practices in the second quintile (20%~40%) of practice statistical frailty have on average 62.7% (standard deviation: 20.0%) patients with missing values on ethnicity and practices in the fourth quintile (60%~80%) of practice statistical frailty also have similar average 64.8% (standard deviation: 23.0%) of patients with missing values on ethnicity. There were no major differences in CVD risk factors and missing levels between practices with high and low statistical frailty. Practices with high/low statistical frailty had comparable means and standard deviations for these characteristics.

Figure 1 shows the relationship between risk factors' dissimilarity between practices (measured by Sáez's metric) and practice statistical frailty. Sáez's metrics for the CVD variables and for their missingness were not related to the statistical frailty of practices (blue lines), indicating that practices with high or low statistical frailty had similar distribution of these risk factors. The same result was found for the overall effects of the coded clinical variables, which suggests that misclassification may not be related to practice statistical frailty. Only a few variables (including Townsend score) were distributed differently between practices with high or low statistical frailty, but there were differences in the patterns between females and males.

Figure 2 (Beeswarm plot) also confirms that there was no visual relationship between practice statistical frailty and most CVD risk factors and their stability metrics (only variables with non-flat trend in Figure 1 are shown). Practices were grouped by percentile of CVD predictors or their stability metrics. The Pearson correlation coefficients between practice frailty and practice characteristics were low. The percentage of smokers had the highest correlations of 0.46 (95% CI: 0.38, 0.54) for females and 0.35 (95%CI: 0.26, 0.44) for males.

Figure 3 shows that there was no variation across practices in the effects of the risk factors on the risk of CVD outcomes, as the fixed effects of QRISK3 linear predictor was near 1 and the variation of random effects on slope among practices was near 0 (0.000111 for females and 0.000302 for males) in the random slope model.

Comparing two reference variations of random slope (0.03 and 0.1), Figure 3 shows that there was almost no random slope, indicating similar associations between predictors and CVD outcome among practices.

As shown in Figure 4, the incorporation of random slopes into the models (i.e., varying effects of the risk factors on CVD between practices) did not increase the accuracy in individual risk prediction. The distributions of the individual risk predictions for patients with a QRISK3 of 10% were comparable between the random slope and the random intercept model. For patients with a QRISK3 predicted 10% risk, the random slope alone would only change the patients' risk by an absolute 0.6% between practices on 97.5% and 2.5% random slope percentile (eFigure3). The effects of variation of random slope on individual patients' risk was small compared with the effects of the random intercept, which could change patient's risk from 10% to a range of 5% and 17%.

**Discussion**

**Key results**

This study found that the observed variation in data quality between general practices did not explain the unmeasured heterogeneity in QRISK3 risk prediction across practices (miscalibration at practice level). Specifically, practices with higher or lower statistical frailty had comparable indicators of data quality, including those based on more innovative techniques (Sáez metric). In addition, the effects of the QRISK3 predictors on CVD risk were comparable across practices despite these differences in data quality, since the random slope models found little variation of the beta coefficients across practices.

**Strength and limitation**

This study was based on a very large patient cohort. It also used the innovative Sáez's metric, which quantifies the distribution-dissimilarity[29]. There are several limitations in this study. We considered several aspects of practice variability that covered important areas identified in literature[4, 11], but there may be other aspects of data quality. The study used Saez metric with distribution-dissimilarity indirectly measured the effects of misclassification for clinical variables. The limitation of our approach was that the multivariate frequencies of clinical variables was evaluated using the Saez metric rather than direct evidence for misclassification. The rationale for our approach was that different coding practices and levels of misclassification are likely to be presented as different levels of frequencies of the variables. Future research might consider a more direct measure of misclassification. Sáez's metric, which measured the CVD risk factors' distribution-dissimilarity among practices, has information loss as it suffers the "curse of dimensionality"[30]. With more practices, there are more dimensions, but this needs to be reduced to estimate summarised statistics resulting in loss of information. Another limitation concerned the estimation of the variation of the random slopes. One thousand bootstrap samples of 40% of the practices were used to estimate this rather than the whole dataset because the current random slope model algorithm[27] has computational difficulty to reach the converge criteria when there is only a small effect on the slopes with greater number of practices. The sensitivity analysis (eFigure 2) shows consistent results of variation of random slope among samples of 20%, 40%, 50% and 60% of total practices suggesting that the variation estimate is accurate.

**Interpretation**

Data quality is an important aspect of practice variability as it influences the performance and generalisability of a model. Damen et al.[4] discussed that data quality limits a model's generalisability and models developed from poor data would generally have poor performance. However, this study shows that although there was a large variation in data quality among practices, it did not affect the accuracy of the risk prediction on individual patients. This indicates data quality among practices were well handled by the data cleaning methods (including multiple imputation for missing values).

Wynants et al.[11] suggested that patient case mix or true variation of association between outcome and predictors might be related to the variation of a model's performance in a heterogeneous setting. Patient case mix was already adjusted in QRISK3[6], and the present study found no random slopes for beta coefficients across practices. This indicates that the effects of QRISK3 predictors on CVD risk were comparable across practices. The comparison between the random slope and random intercept models found that the effects of practice variability on individual patients was fully explained by the random intercept, i.e. the unmeasured heterogeneity in CVD incidence between practices and deviation from the baseline hazard.

A recent study[31] found that the addition of another risk factor (standard deviation of blood pressure) to QRISK3 did not improve model performance despite it being significantly related to CVD. Previous studies[9] discovered models with similar discrimination and calibration could predict the same patient differently using the current model's predictors. Therneau[32] demonstrated an example that the effects of random intercept could come from unknown covariates missed by a model. This study found that unexplained heterogeneity at practice level cannot be resolved using current measured risk factors. Therefore, this study supports the conclusion from Damen et al.[4] that current CVD models lack information on other important CVD risk factors, e.g. those that better measure the heterogeneity in incidence between different areas.

**Implications for research and practice**

This study found that variation between practices in data quality and effects of CVD predictors were not associated with the considerable heterogeneity in CVD incidence. This suggests that a further study might focus on determining whether the CVD risk prediction models can be extended with new risk factors from patient level CVD risk factors (e.g. biomarkers) or practice level, which could reduce the unmeasured heterogeneity in CVD incidence across practices. Further research could consider more individual level based methods, such as a Bayesian clinical reasoning model[33] and machine learning models[34], as this study and other findings[18, 35, 36] show that Cox models with similar conventional model performance metrics (C-stat[37] and calibration) could predict inconsistent risk to the same patients. Alternatively, new statistics might be required to measure a population-based model performance on an individual level.

In conclusion, the considerable unmeasured heterogeneity in CVD incidence between practices was not explained by variations in data quality or effects of risk factors. QRISK3 risk prediction should be supplemented with clinical judgement and evidence of additional risk factors.

**Funding**

This study was funded by China Scholarship Council (PhD studentship of Yan Li).

**Acknowledgements**

This study is based on data from the Clinical Practice Research Datalink obtained under license from the UK Medicines and Healthcare products Regulatory Agency. The protocol for this work was approved by the independent scientific advisory committee for Clinical Practice Research Datalink research (No 17_125RMn2). The data are provided by patients and collected by the NHS as part of their care and support. The Office for National Statistics (ONS) is the provider of the ONS Data contained within the CPRD Data. Hospital Episode Data and the ONS Data Copyright © (2014), are re-used with the permission of The Health & Social Care Information Centre. All rights reserved. The interpretation and conclusions contained in this study are those of the authors alone. There are no conflicts of interest among the authors.

**Summary points**

**What was already known**

- Risk prediction tools based on routinely collected data are used by clinicians to predict a 10-year CVD risk for individual patients.
- A previous study found that there was considerable variability between clinical sites in the robustness of individual risk predictions. This heterogeneity in incidence between sites is not incorporated into current risk prediction approaches.

**What this study has added**

- There was substantial heterogeneity between practices in the incidence of cardiovascular disease (CVD) which was not explained by a commonly used risk prediction model (QRISK3).
- Data quality, as measured by probabilistic indicators based on information theory and geometry, varied substantially between clinical sites.
- This study adds that this unmeasured heterogeneity in CVD incidence was not explained by variations in data quality or effects of risk factors between clinical sites.

**Table 1: Predicted CVD risks in random intercept models (for patients with QRISK3 predicted risk of 10%) stratified into quintiles based on the level of differences between these predictions**

| Quintile of practice frailty | Number of practices | Frailty | Predicted CVD risk with random intercept model (%) |
|---|---|---|---|
| **Female** | | | |
| 0~20% | 78 | 0.7~ 0.9 | 7.1~ 9.0 |
| 20~40% | 78 | 0.9~ 1.0 | 9.0~ 10.0 |
| 40~60% | 79 | 1.0~ 1.0 | 10.0~ 10.0 |
| 60~80% | 78 | 1.0~ 1.1 | 10.0~ 10.9 |
| 80~100% | 79 | 1.1~ 1.7 | 10.9~ 16.4 |
| **Male** | | | |
| 0~20% | 78 | 0.6~ 0.9 | 6.1~ 9.0 |
| 20~40% | 78 | 0.9~ 1.0 | 9.0~ 10.0 |
| 40~60% | 79 | 1.0~ 1.0 | 10.0~ 10.0 |
| 60~80% | 78 | 1.0~ 1.1 | 10.0~ 10.9 |
| 80~100% | 79 | 1.1~ 1.6 | 10.9~ 15.5 |

**Table 2: Characteristics of the practices stratified by different quintiles of statistical frailty**

| | Male | | | | | Female | | | | |
|---|---|---|---|---|---|---|---|---|---|---|
| | mean (SD)) | | | | | (mean (SD)) | | | | |
| | Frailty (0~20%) ( 0.7 ~ 0.9) | Frailty (20~40%) ( 0.9 ~ 1.0) | Frailty (40~60%) ( 1.0 ~ 1.0) | Frailty (60~80%) ( 1.0 ~ 1.1) | Frailty (80~100%) ( 1.1 ~ 1.7) | Frailty (0~20%) ( 0.6 ~ 0.9) | Frailty (20~40%) ( 0.9 ~ 1.0) | Frailty (40~60%) ( 1.0 ~ 1.0) | Frailty (60~80%) ( 1.0 ~ 1.1) | Frailty (80~100%) ( 1.1 ~ 1.6) |
| **General characteristics of practices** | | | | | | | | | | |
| Average number of CVD events in 10 years within practice strata by gender | 84.5 (58.5) | 133.1 (89.3) | 142.0 (92.1) | 181.2 (101.7) | 182.4 (87.0) | 89.0 (57.6) | 104.0 (86.3) | 122.7 (91.5) | 143.4 (80.3) | 159.5 (75.8) |
| Average age within practice | 43.6 (2.9) | 44.5 (3.0) | 44.2 (3.1) | 44.7 (2.4) | 44.2 (2.0) | 44.9 (4.0) | 46.0 (3.8) | 45.0 (3.6) | 46.2 (2.7) | 45.6 (2.6) |
| Average number of patients within practice strata by gender at index date | 4528.0 (2543.6) | 4680.9 (2737.6) | 4330.1 (2392.9) | 4930.5 (2723.2) | 4112.1 (1825.2) | 5415.8 (2805.9) | 4590.3 (2951.9) | 4578.5 (2767.0) | 4819.0 (2375.8) | 4341.2 (2012.4) |
| Number of practices | 78 | 78 | 79 | 78 | 79 | 78 | 78 | 79 | 78 | 79 |
| **CVD risk factors** | | | | | | | | | | |
| % patients with alcohol abuse | 1.5 (0.8) | 1.6 (1.0) | 1.8 (1.7) | 2.0 (2.0) | 2.4 (1.3) | 0.7 (0.4) | 1.0 (1.7) | 0.9 (0.7) | 0.8 (0.4) | 1.1 (0.7) |
| % patients with anxiety | 8.9 (3.5) | 8.9 (3.0) | 9.7 (3.7) | 10.9 (3.4) | 12.0 (5.2) | 15.5 (5.9) | 15.9 (5.9) | 17.2 (6.7) | 17.4 (6.3) | 20.1 (7.6) |
| % patients with HIV | 0.1 (0.2) | 0.1 (0.1) | 0.1 (0.1) | 0.1 (0.1) | 0.1 (0.1) | 0.1 (0.1) | 0.1 (0.1) | 0.1 (0.2) | 0.1 (0.1) | 0.0 (0.1) |
| % patients with left ventricular hypertrophy | 0.2 (0.1) | 0.2 (0.1) | 0.3 (0.2) | 0.3 (0.1) | 0.3 (0.2) | 0.2 (0.1) | 0.2 (0.1) | 0.2 (0.2) | 0.2 (0.1) | 0.2 (0.2) |
| % patients with atrial fibrillation | 0.8 (0.4) | 0.9 (0.4) | 0.8 (0.4) | 0.9 (0.3) | 0.7 (0.3) | 0.6 (0.3) | 0.7 (0.3) | 0.6 (0.3) | 0.7 (0.3) | 0.6 (0.3) |
| % patients on atypical antipsychotic medication | 0.4 (0.3) | 0.4 (0.3) | 0.4 (0.3) | 0.4 (0.2) | 0.5 (0.2) | 0.4 (0.2) | 0.4 (0.2) | 0.4 (0.3) | 0.4 (0.2) | 0.5 (0.2) |
| % patients with chronic kidney disease (stage 3, 4 or 5) | 0.8 (0.4) | 0.8 (1.0) | 0.8 (0.5) | 0.7 (0.4) | 0.6 (0.3) | 1.3 (0.9) | 1.5 (1.0) | 1.5 (2.2) | 1.3 (0.8) | 1.1 (0.6) |
| % patients on regular steroid tablets | 0.1 (0.1) | 0.1 (0.1) | 0.1 (0.1) | 0.1 (0.1) | 0.1 (0.1) | 0.1 (0.1) | 0.1 (0.1) | 0.1 (0.1) | 0.1 (0.1) | 0.1 (0.1) |
| % patients with angina or heart attack in a 1st degree relative < 60 | 3.5 (2.5) | 3.5 (3.1) | 3.2 (2.2) | 2.9 (2.8) | 2.8 (2.2) | 4.5 (3.8) | 4.3 (3.5) | 4.0 (2.7) | 4.0 (4.1) | 3.2 (2.4) |
| % patients on blood pressure treatment | 5.5 (2.0) | 5.9 (1.6) | 6.0 (1.7) | 6.0 (1.5) | 5.6 (1.5) | 7.4 (2.7) | 8.2 (2.6) | 7.6 (2.0) | 8.2 (1.8) | 7.8 (2.2) |
| % patients with migraines | 3.3 (1.3) | 3.3 (1.2) | 3.6 (1.3) | 3.7 (1.2) | 3.6 (1.5) | 8.7 (3.0) | 8.5 (2.7) | 9.6 (2.9) | 9.3 (2.8) | 9.7 (3.6) |
| % patients with rheumatoid arthritis | 0.3 (0.2) | 0.3 (0.2) | 0.3 (0.1) | 0.4 (0.1) | 0.4 (0.2) | 0.8 (0.3) | 0.9 (0.3) | 0.9 (0.4) | 0.9 (0.3) | 0.9 (0.3) |
| % patients with severe mental illness | 4.5 (2.7) | 5.3 (3.0) | 4.8 (2.7) | 6.2 (3.0) | 6.4 (3.3) | 8.1 (5.2) | 8.8 (4.9) | 10.4 (7.0) | 11.1 (6.1) | 12.0 (6.5) |

|  | Male | | | | | | Female | | | | |
|---|---|---|---|---|---|---|---|---|---|---|---|
|  | mean (SD)) | | | | | | (mean (SD)) | | | | |
|  | Frailty (0~20%) ( 0.7 ~ 0.9) | Frailty (20~40%) ( 0.9 ~ 1.0) | Frailty (40~60%) ( 1.0 ~ 1.0) | Frailty (60~80%) ( 1.0 ~ 1.1) | Frailty (80~100%) ( 1.1 ~ 1.7) | | Frailty (0~20%) ( 0.6 ~ 0.9) | Frailty (20~40%) ( 0.9 ~ 1.0) | Frailty (40~60%) ( 1.0 ~ 1.0) | Frailty (60~80%) ( 1.0 ~ 1.1) | Frailty (80~100%) ( 1.1 ~ 1.6) |
| % patients with Systemic Lupus Erythematosus | 0.0 (0.0) | 0.0 (0.0) | 0.0 (0.0) | 0.0 (0.0) | 0.0 (0.0) | | 0.1 (0.1) | 0.1 (0.1) | 0.1 (0.1) | 0.1 (0.1) | 0.1 (0.1) |
| **SBP** | | | | | | | | | | | |
| Average SBP within practice | 130.1 (2.6) | 130.4 (2.8) | 130.6 (3.2) | 130.8 (2.6) | 130.6 (2.5) | | 123.4 (3.2) | 124.4 (3.2) | 123.8 (3.2) | 124.8 (2.7) | 125.0 (2.5) |
| % patients with missing SBP | 36.2 (9.8) | 34.9 (8.4) | 36.7 (8.9) | 37.5 (8.4) | 38.3 (9.4) | | 14.7 (5.8) | 13.7 (5.8) | 13.9 (5.3) | 14.6 (6.6) | 16.5 (6.5) |
| **BMI** | | | | | | | | | | | |
| Average BMI when recorded | 26.5 (0.7) | 26.5 (0.6) | 26.7 (0.5) | 26.7 (0.6) | 26.7 (0.5) | | 25.7 (1.1) | 26.1 (1.0) | 26.2 (0.9) | 26.3 (0.6) | 26.6 (0.7) |
| % patients with missing BMI | 45.6 (12.5) | 47.3 (11.2) | 47.2 (13.3) | 50.2 (10.7) | 50.0 (12.1) | | 30.4 (11.7) | 29.1 (13.6) | 28.9 (11.8) | 31.3 (11.7) | 33.6 (12.9) |
| **Cholesterol/HDL ratio** | | | | | | | | | | | |
| Average Cholesterol/HDL ratio within practice | 4.3 (0.2) | 4.4 (0.2) | 4.4 (0.3) | 4.4 (0.2) | 4.4 (0.2) | | 3.6 (0.2) | 3.6 (0.2) | 3.6 (0.3) | 3.7 (0.2) | 3.8 (0.2) |
| % patients with missing Cholesterol/HDL ratio | 65.6 (10.4) | 66.3 (8.8) | 64.8 (8.9) | 69.0 (9.4) | 65.5 (8.1) | | 63.3 (10.4) | 60.1 (11.2) | 61.8 (10.7) | 64.9 (11.0) | 62.7 (11.4) |
| **Smoking** | | | | | | | | | | | |
| % current-smokers | 32.3 (8.0) | 32.9 (6.1) | 33.5 (6.1) | 36.4 (6.9) | 38.8 (6.7) | | 22.1 (5.5) | 24.3 (6.5) | 24.4 (7.4) | 26.5 (5.4) | 31.0 (6.7) |
| % patients with missing smoking status | 26.2 (9.4) | 28.6 (8.2) | 27.7 (9.7) | 31.7 (8.8) | 33.1 (8.1) | | 17.4 (7.5) | 17.8 (8.4) | 16.9 (8.4) | 21.0 (8.3) | 22.9 (8.8) |
| **Diabetes** | | | | | | | | | | | |
| % patients with type 1 diabetes | 0.2 (0.1) | 0.2 (0.1) | 0.2 (0.1) | 0.2 (0.1) | 0.3 (0.1) | | 0.2 (0.1) | 0.2 (0.1) | 0.2 (0.1) | 0.2 (0.1) | 0.2 (0.1) |
| % patients with type 2 diabetes | 1.2 (0.5) | 1.3 (0.4) | 1.5 (0.5) | 1.6 (0.4) | 1.6 (0.4) | | 1.0 (0.4) | 1.1 (0.4) | 1.1 (0.4) | 1.3 (0.4) | 1.3 (0.5) |
| **Ethnicity** | | | | | | | | | | | |
| % white patients | 83.0 (16.3) | 87.4 (15.5) | 83.2 (20.0) | 89.7 (11.0) | 90.1 (12.9) | | 84.6 (15.9) | 86.4 (16.9) | 83.9 (18.1) | 88.9 (12.1) | 90.3 (11.8) |
| % patients with missing ethnicity | 55.7 (21.7) | 62.7 (20.0) | 57.8 (26.6) | 64.8 (23.0) | 62.1 (22.4) | | 54.2 (23.0) | 54.9 (26.3) | 53.2 (25.1) | 61.4 (23.1) | 58.9 (24.7) |
| **Townsend (Socioeconomic Status)** | | | | | | | | | | | |
| % patients with Townsend score 5 (the most deprived) | 16.2 (25.8) | 11.9 (21.0) | 15.1 (21.9) | 14.0 (19.1) | 24.5 (20.1) | | 10.0 (19.5) | 14.9 (22.4) | 17.2 (24.0) | 11.2 (16.4) | 25.9 (21.6) |
| % patients with Townsend score missing | 0.1 (0.1) | 0.1 (0.1) | 0.1 (0.1) | 0.1 (0.1) | 0.2 (1.4) | | 0.1 (0.1) | 0.1 (0.1) | 0.1 (0.1) | 0.1 (0.1) | 0.2 (1.4) |

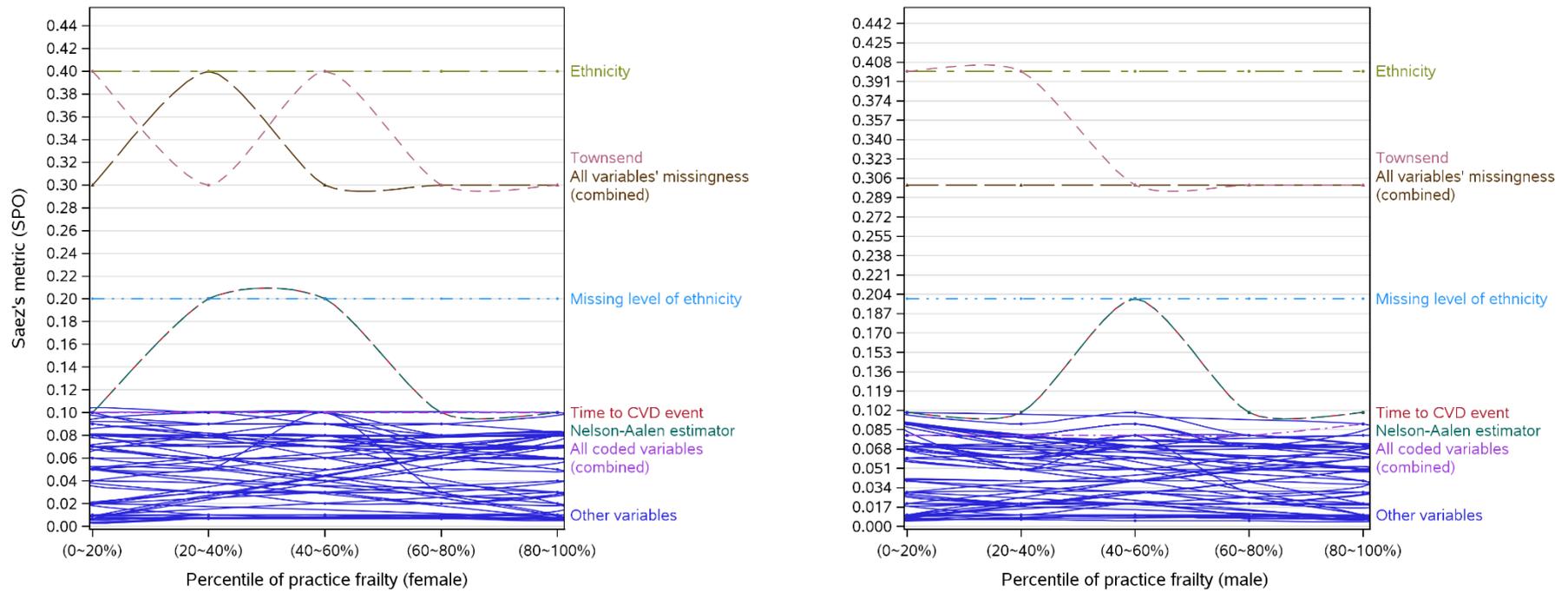

**Figure 1 Relationship between quintiles of statistical frailty in practices and the stability metrics for QRISK3 CVD predictors and level of missingness**

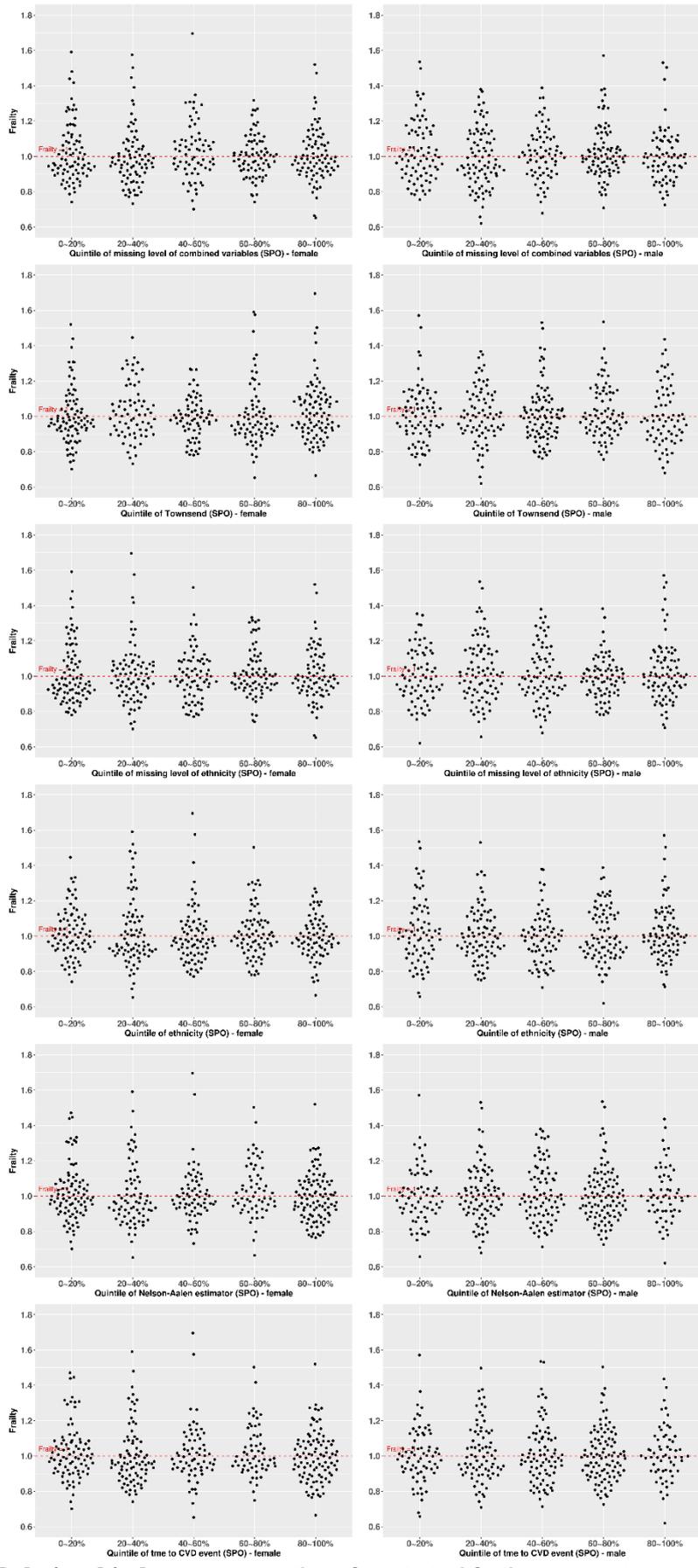

**Figure 2. Relationship between quintiles of statistical frailty in practices and CVD risk predictors and their stability metrics (SPO) - Beeswarm plot**

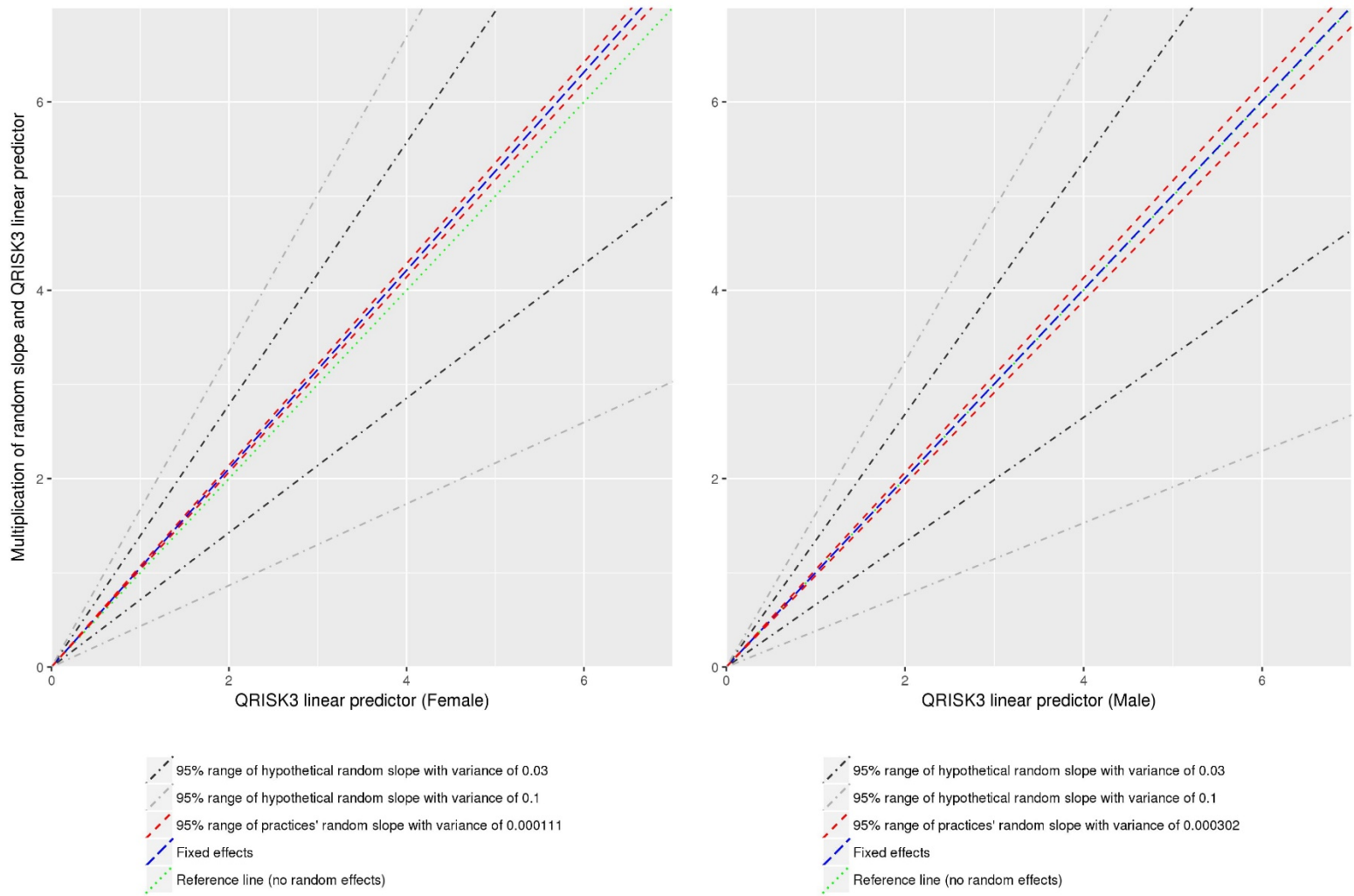

**Figure 3 Effects of the variability between practices of the QRISK3 linear predictor (random slope)**

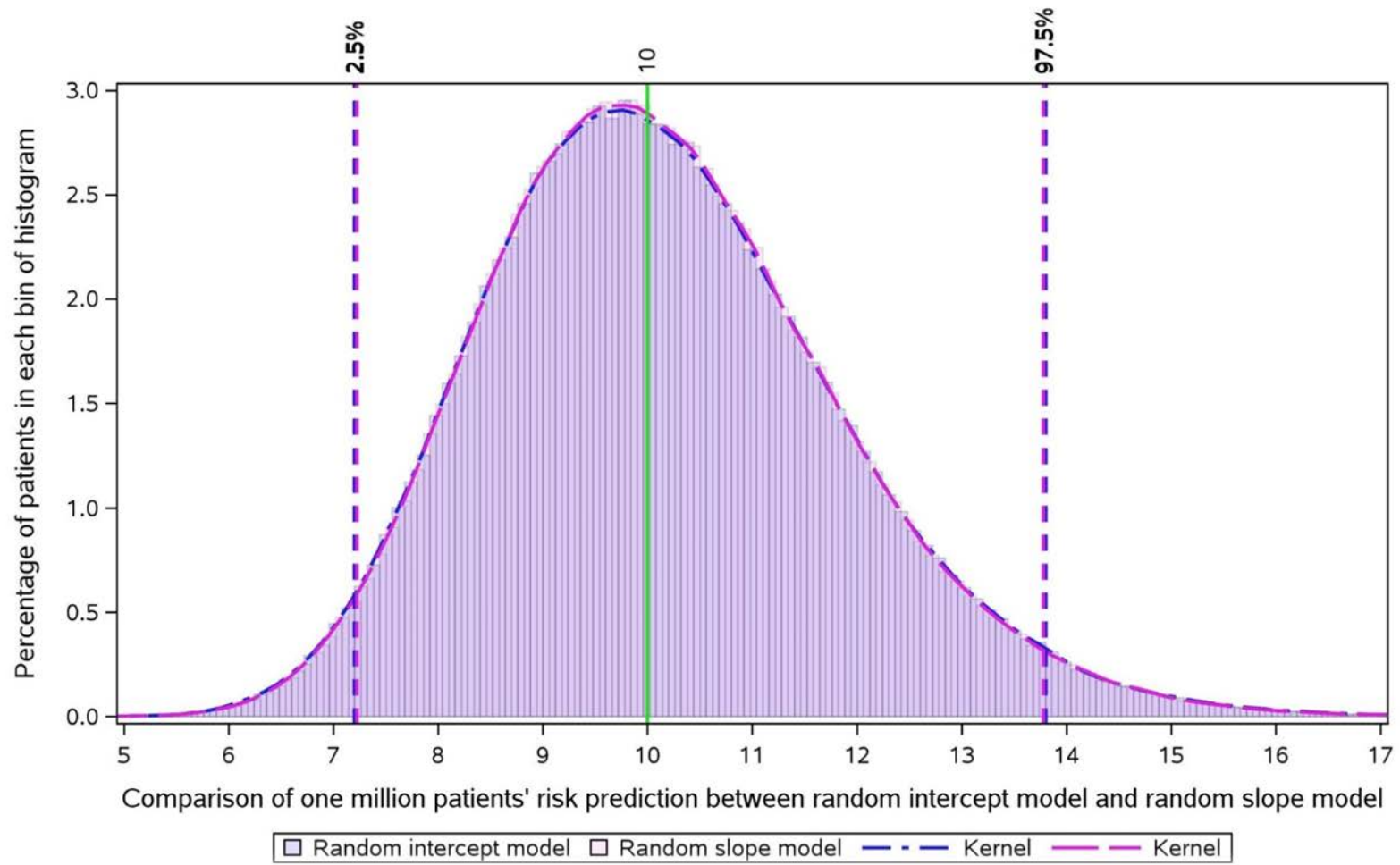

**Figure 4 Comparison of the CVD risk predictions between the random intercept and slope models for patients with a QRISK3 risk of 10% (in a cohort of one million patients with 50% males and 50% females)**

**Supplementary Online Content**

Title: Examining the impact of data quality and completeness of electronic health records on predictions of patients' risks of cardiovascular disease
Journal title: International Journal of Medical Informatics
Doi: https://doi.org/10.1016/j.ijmedinf.2019.104033

**eAppendix 1.** Interpretation of appendix tables and figures.

**eAppendix 2.** Saez's metric of distribution-dissimilarity

**References**

**eTable 1.** Stability metrics of all QRISK3 CVD predictors and their missing level on practice level

**eFigure 1**. Relationship between practice frailty and CVD risk predictors and their stability metrics (Beeswarm plot).

**eFigure 2-1**. Effects of practice variability on QRISK3 linear predictor (random slope) (20% of overall CPRD practices)

**eFigure 2-2**. Effects of practice variability on QRISK3 linear predictor (random slope) (50% of overall CPRD practices)

**eFigure 2-3**. Effects of practice variability on QRISK3 linear predictor (random slope) (60% of overall CPRD practices)

**eFigure 3.** Difference of individual patients' prediction between practice with 2.5% random slope and 97.5% slope and a random selected fixed random intercept

**eAppendix 1.** Interpretation of appendix tables and figures.

eTable1 summarises the distribution-dissimilarity of all CVD risk factors and their missingness using Sáez's proposed metrics global probabilistic deviation (GPD) and source probabilistic outlyingness (SPO). GPD measures risk factors' overall outlyingness of practices, which ranges from 0 to 1 and the closer to 1 means there is more variation of practices. SPO measures the latent distance of risk factors' distribution between one practice to the average of overall practice, which also ranges from 0 to 1 and closer to 1 means the practice is more far away from the average. The table shows part of CVD risk factors (e.g. Rheumatoid arthritis) among practices are very stable, which means they have similar distribution among practices. Other variables, such as ethnicity, Townsend and missing level of ethnicity, were unstable among practices, which means the distribution of these variables has substantial variation among practices.

eFigure1 visualises the distribution-dissimilarity of all CVD risk factors and their missingness using percentile of Sáez's proposed metrics SPO strata by gender. The result is consistent with eTable1, as part of risk factors are very stable among practices (blue lines), and other variables such as Townsend and ethnicity has substantial variation among practices.

Preliminary analysis showed that the variation of random slopes of full CPRD practices was too small and current statistical package would take very long to calculate results for full practices, so the study estimated practices' variation of random slope by averaging practices' variation of random slope of 1000 random samples (each sample contained 40% practices of all CPRD practices). Sensitive analysis (eFigure 2) shows that there is no difference of the average variation of random slope among different sample size of practice (20%, 40%, 50% and 60% percent of full CPRD practices). All of them shows that there is no variation of random slope among practices, which suggests all practices have similar association between predictors and outcome. This also suggests all of samples are a representative sample of CPRD practices, just as CPRD is a representative sample of the whole UK practices.

eFigure3 shows that for patients with a QRISK3 predicted 10% risk, random slope alone would only change the patients' risk by absolute 0.6% between practices on 97.5% and 2.5% random slope percentile. The effects of variation of random slope on individual patients' risk is small comparing to that random intercept could change patient's risk from 10% a range of 5% and 17%.The effects of random slope on individual patients however increases with the increase of patients predicted risk by QRISK3, but it would not affect patients' classification at most of the time. For example, although patients' risk could change about 2% when they have 25%, but the patients would still be prescribed statin after change. Also, the larger patients' predicted risk by QRISK3 means the larger linear predictor which then enlarges the effects of random slope through exponential function from Cox model. Consider following empirical example which compares exp (10*1.05) - exp (10*1.01) = 11972.49 to exp (20*1.05) – exp (20*1.01) = 726233627. We think 20 is a linear predictor for patients with a very high risk, and 10 is for a patient with low risk. Ignoring random intercept here and think 1.01 is a sum of fixed and random slope. Say the fix slope is 1, and 97.5 random slope is 0.05 and 2.5% random slope is 0.01. We can see the larger linear predictor enlarges the differences because of exponential function here.

**eAppendix 2.** Technical details of Sáez's metric of distribution-dissimilarity

Sáez proposed non-parametric information theory metrics to quantify the distribution-dissimilarity of single or multiple variables among different practice (sites) [1]. Saez quantified the distribution-dissimilarity of variables using Jensen–Shannon divergence (JSD) [2]. JSD (ranges from 0 to 1) calculates an information distance (divergence) of the variable's probability distribution in different practices (sites), which measures the distribution-dissimilarity of a variable among different practices (sites). Once the information distance of a variable between all pair of practices were acquired, Euclidean embedding [3] and simplex [4] theory could be used to construct a coordinate for each practice based on the information distance among them. Based on the geometry theory, the coordinate of a latent center (centroid [4]) could be calculated by averaging all practices' coordinates. The centroid represents a latent average distribution of the variable among all practices, so the information distance between one practice to the centroid quantifies the dissimilarity of variable's distribution of one practice to the overall average. By standardising this distance (so the information distance of different variables is comparable), Sáez proposed source probabilistic outlyingness (SPO) metric, which ranges from 0 to 1 and the higher means the site is more far away from centroid, to quantify the distribution-dissimilarity of variable from one practice to the overall average. Sáez also proposed global probabilistic deviation (GPD), also ranges from 0 to 1 and the higher means the more variation of the variable among practices, to quantify variable's the overall distribution stability among practices [1]. For multiple variables, dimension reduction method such as principle component analysis (PCA) [3] and factor analysis [3], could be used to construct three or four independent principle components to represent the overall variation of multiple variables. Joint probabilities could be calculated using these principle components' distribution, and then JSD of joint probabilities could be calculated to represent the combined distribution-dissimilarity of multiple variables among practices.

**eTable 1. Stability metrics of all QRISK3 CVD predictors and their missing level on practice level**

|  | Male | | | | | | Female | | | | | |
|---|---|---|---|---|---|---|---|---|---|---|---|---|
|  | GPD | 5th | 25th | 50th | 75th | 95th | GPD | 5th | 25th | 50th | 75th | 95th |
| **CVD risk factors (distribution dissimilarity to the overall practice average)** | | | | | | | | | | | | |
| Atrial fibrillation | 0.02 | 0.00 | 0.01 | 0.01 | 0.02 | 0.04 | 0.02 | 0.00 | 0.01 | 0.01 | 0.02 | 0.03 |
| Whether patients on atypical antipsychotic medication | 0.02 | 0.00 | 0.01 | 0.01 | 0.02 | 0.03 | 0.02 | 0.00 | 0.00 | 0.01 | 0.02 | 0.03 |
| Chronic kidney disease (stage 3, 4 or 5) | 0.02 | 0.00 | 0.01 | 0.01 | 0.02 | 0.04 | 0.03 | 0.00 | 0.01 | 0.02 | 0.03 | 0.05 |
| CVD censors | 0.08 | 0.02 | 0.02 | 0.03 | 0.05 | 0.11 | 0.08 | 0.02 | 0.02 | 0.03 | 0.05 | 0.10 |
| Time to CVD event | 0.19 | 0.05 | 0.07 | 0.10 | 0.14 | 0.40 | 0.20 | 0.05 | 0.08 | 0.11 | 0.15 | 0.42 |
| Cholesterol | 0.08 | 0.02 | 0.03 | 0.05 | 0.07 | 0.11 | 0.10 | 0.02 | 0.04 | 0.06 | 0.09 | 0.17 |
| Regular steroid tablets | 0.01 | 0.00 | 0.00 | 0.01 | 0.01 | 0.02 | 0.01 | 0.00 | 0.00 | 0.01 | 0.01 | 0.02 |
| Erectile dysfunction | 0.03 | 0.00 | 0.01 | 0.02 | 0.03 | 0.06 | 0.01 | 0.00 | 0.00 | 0.01 | 0.01 | 0.02 |
| Angina or heart attack in a 1st degree relative < 60 | 0.06 | 0.01 | 0.02 | 0.03 | 0.06 | 0.10 | 0.07 | 0.01 | 0.02 | 0.04 | 0.06 | 0.11 |
| HDL | 0.11 | 0.03 | 0.06 | 0.07 | 0.10 | 0.15 | 0.12 | 0.03 | 0.05 | 0.08 | 0.11 | 0.17 |
| Blood pressure treatment | 0.03 | 0.00 | 0.01 | 0.02 | 0.03 | 0.06 | 0.04 | 0.00 | 0.01 | 0.02 | 0.04 | 0.07 |
| Migraines | 0.03 | 0.00 | 0.01 | 0.02 | 0.03 | 0.06 | 0.05 | 0.00 | 0.02 | 0.03 | 0.05 | 0.10 |
| Nelson-Aalen estimator | 0.18 | 0.05 | 0.07 | 0.10 | 0.14 | 0.36 | 0.19 | 0.05 | 0.08 | 0.11 | 0.14 | 0.36 |
| Rheumatoid arthritis | 0.01 | 0.00 | 0.00 | 0.01 | 0.01 | 0.02 | 0.02 | 0.00 | 0.00 | 0.01 | 0.01 | 0.03 |
| Systemic Lupus Erythematosus | 0.01 | 0.00 | 0.00 | 0.00 | 0.01 | 0.01 | 0.01 | 0.00 | 0.00 | 0.01 | 0.01 | 0.02 |
| Severe mental illness (this includes schizophrenia, bipolar disorder and moderate/severe depression) | 0.07 | 0.01 | 0.02 | 0.04 | 0.07 | 0.10 | 0.10 | 0.01 | 0.04 | 0.07 | 0.10 | 0.15 |
| Type 1 diabetes | 0.01 | 0.00 | 0.00 | 0.01 | 0.01 | 0.02 | 0.01 | 0.00 | 0.00 | 0.01 | 0.01 | 0.02 |

|  | Male | | | | | | Female | | | | | |
| --- | --- | --- | --- | --- | --- | --- | --- | --- | --- | --- | --- | --- |
|  | GPD | 5th | 25th | 50th | 75th | 95th | GPD | 5th | 25th | 50th | 75th | 95th |
| Type 2 diabetes | 0.02 | 0.00 | 0.01 | 0.01 | 0.02 | 0.03 | 0.02 | 0.00 | 0.01 | 0.01 | 0.02 | 0.04 |
| Age | 0.11 | 0.02 | 0.04 | 0.07 | 0.10 | 0.17 | 0.12 | 0.03 | 0.05 | 0.07 | 0.12 | 0.20 |
| BMI | 0.09 | 0.02 | 0.04 | 0.06 | 0.08 | 0.12 | 0.09 | 0.02 | 0.04 | 0.06 | 0.08 | 0.14 |
| Cholesterol and HDL | 0.11 | 0.03 | 0.05 | 0.07 | 0.09 | 0.15 | 0.12 | 0.03 | 0.05 | 0.07 | 0.11 | 0.17 |
| Ethnicity | 0.57 | 0.27 | 0.32 | 0.38 | 0.47 | 0.63 | 0.61 | 0.29 | 0.34 | 0.40 | 0.50 | 0.63 |
| SBP | 0.13 | 0.03 | 0.06 | 0.08 | 0.11 | 0.18 | 0.12 | 0.03 | 0.05 | 0.08 | 0.10 | 0.17 |
| Standard deviation of SBP | 0.09 | 0.02 | 0.04 | 0.06 | 0.08 | 0.13 | 0.08 | 0.02 | 0.04 | 0.05 | 0.07 | 0.12 |
| Smoking | 0.10 | 0.02 | 0.04 | 0.06 | 0.09 | 0.16 | 0.11 | 0.02 | 0.04 | 0.06 | 0.10 | 0.19 |
| Townsend | 0.48 | 0.17 | 0.24 | 0.32 | 0.43 | 0.58 | 0.48 | 0.17 | 0.24 | 0.31 | 0.43 | 0.56 |
| **Missing level (distribution dissimilarity to the overall practice average)** | | | | | | | | | | | | |
| Missing level of Cholesterol | 0.07 | 0.01 | 0.02 | 0.04 | 0.08 | 0.14 | 0.09 | 0.01 | 0.03 | 0.05 | 0.09 | 0.16 |
| Missing level of HDL | 0.09 | 0.01 | 0.03 | 0.05 | 0.09 | 0.18 | 0.11 | 0.01 | 0.03 | 0.06 | 0.11 | 0.21 |
| Missing level of Nelson-Aalen estimator | 0.04 | 0.00 | 0.01 | 0.02 | 0.03 | 0.06 | 0.03 | 0.00 | 0.01 | 0.02 | 0.03 | 0.06 |
| Missing level of BMI | 0.12 | 0.01 | 0.03 | 0.07 | 0.12 | 0.22 | 0.13 | 0.01 | 0.04 | 0.08 | 0.13 | 0.24 |
| Missing level of ratio of Cholesterol and HDL | 0.09 | 0.01 | 0.03 | 0.05 | 0.09 | 0.18 | 0.11 | 0.01 | 0.03 | 0.06 | 0.11 | 0.21 |
| Missing level of ethnicity | 0.26 | 0.04 | 0.10 | 0.16 | 0.24 | 0.43 | 0.28 | 0.05 | 0.11 | 0.18 | 0.25 | 0.44 |
| Missing level of SBP | 0.09 | 0.01 | 0.02 | 0.05 | 0.09 | 0.17 | 0.08 | 0.01 | 0.02 | 0.05 | 0.08 | 0.14 |
| Missing level of standard deviation of SBP | 0.07 | 0.00 | 0.02 | 0.04 | 0.07 | 0.14 | 0.08 | 0.01 | 0.02 | 0.05 | 0.08 | 0.15 |
| Missing level of smoking | 0.10 | 0.01 | 0.03 | 0.06 | 0.10 | 0.18 | 0.11 | 0.01 | 0.03 | 0.06 | 0.11 | 0.19 |
| Missing level of townsend | 0.02 | 0.00 | 0.00 | 0.01 | 0.02 | 0.02 | 0.01 | 0.00 | 0.00 | 0.01 | 0.02 | 0.02 |

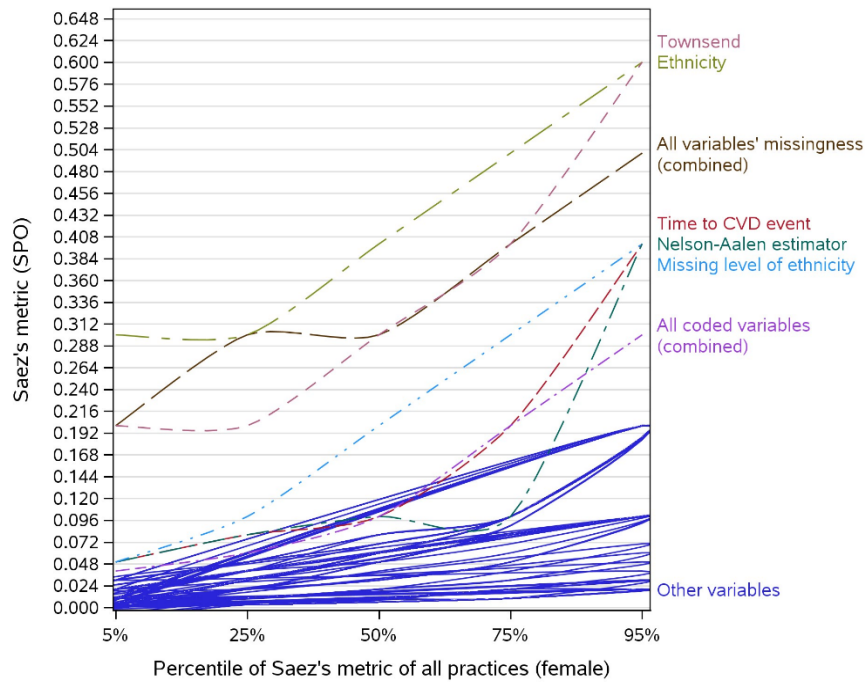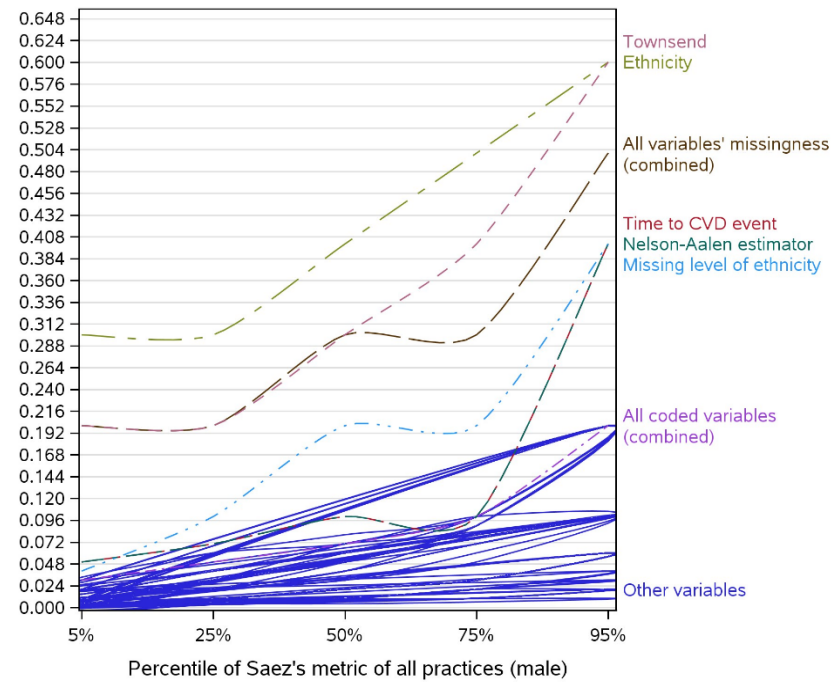

**eFigure 1 Stability metrics of all QRISK3 CVD predictors and their missing level on practice level**

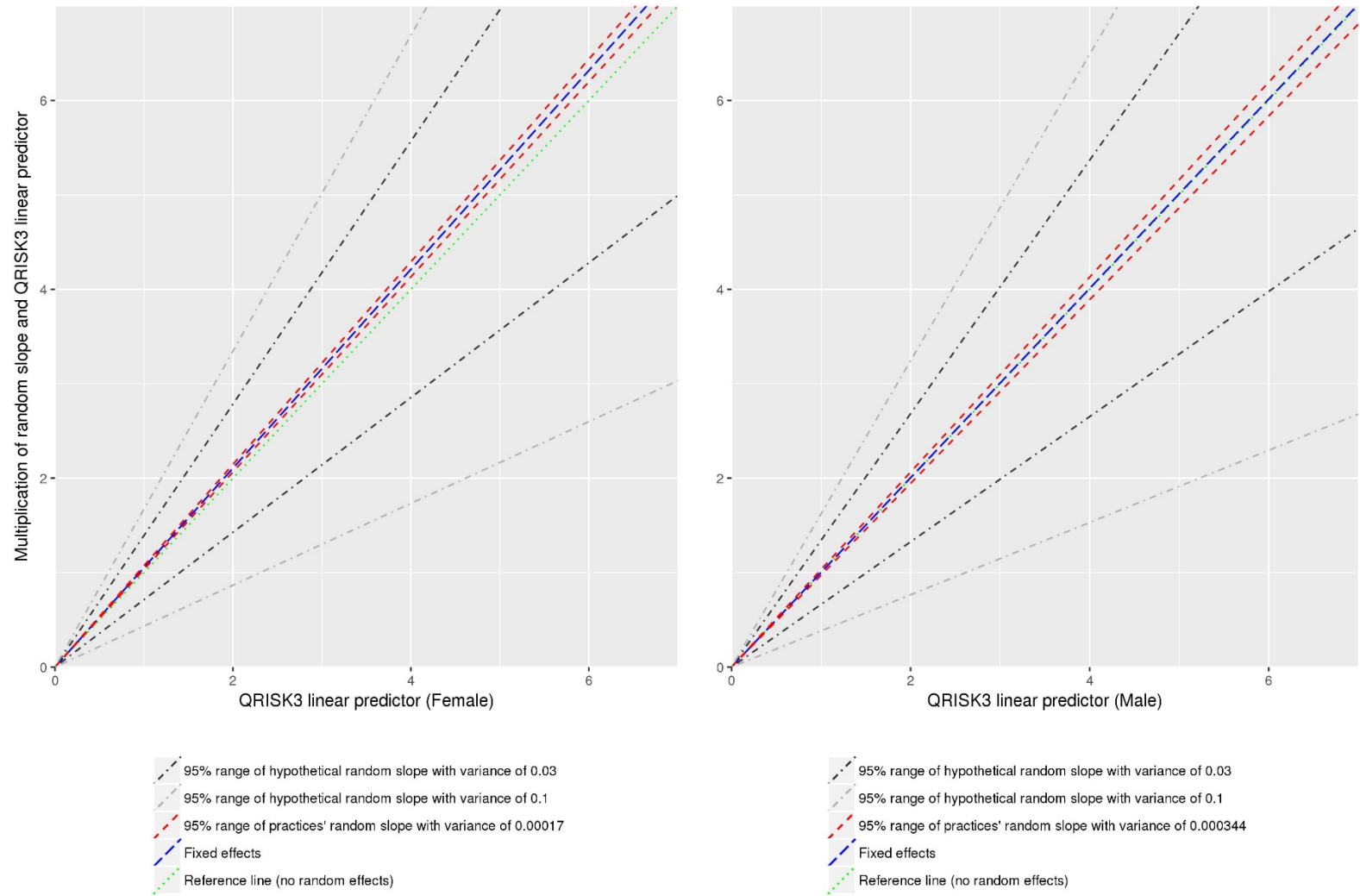

**eFigure 2-1**. Effects of practice variability on QRISK3 linear predictor (random slope) (20% of overall CPRD practices)

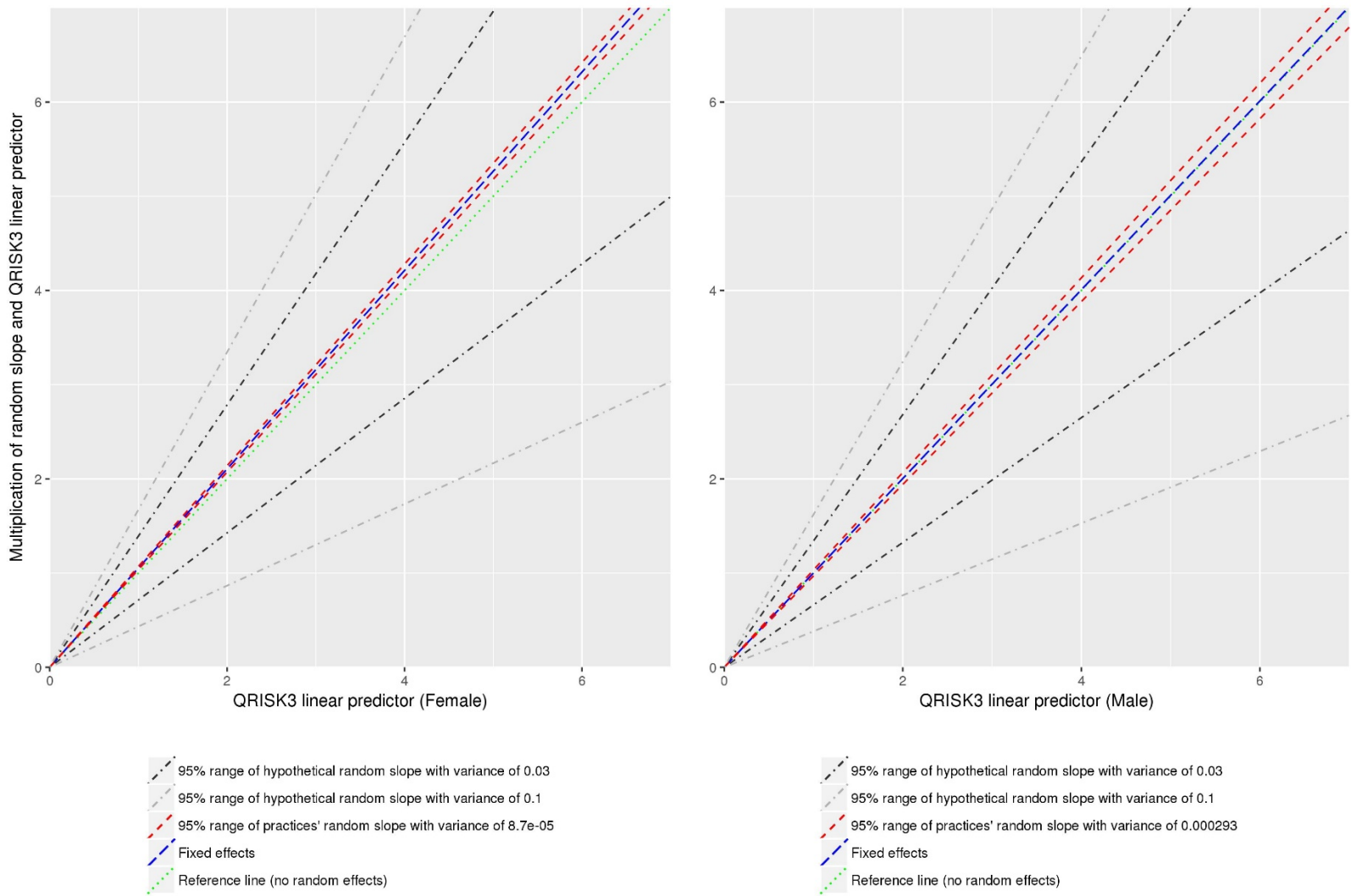

**eFigure 2-2.** Effects of practice variability on QRISK3 linear predictor (random slope) (50% of overall CPRD practices)

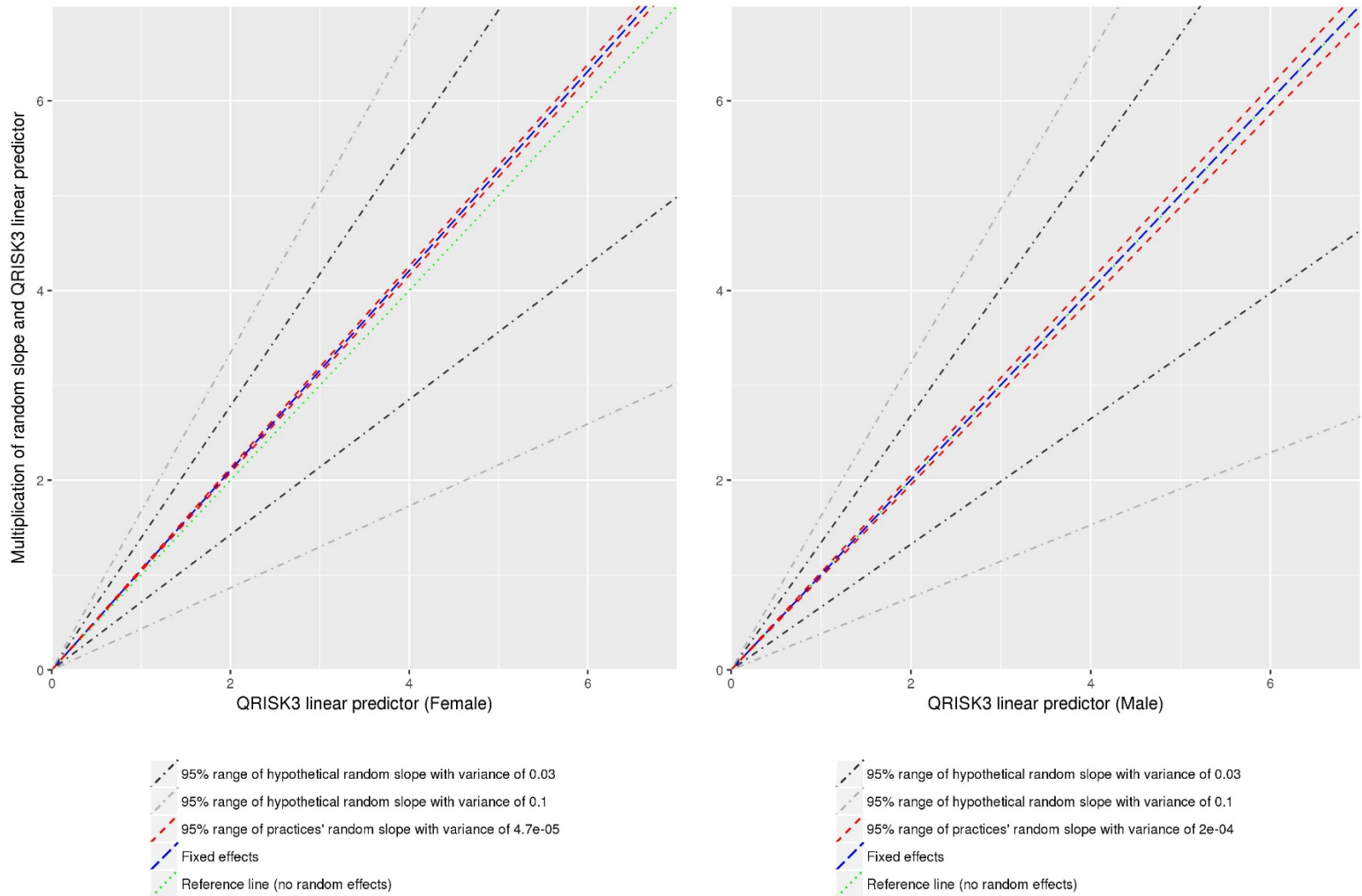

**eFigure 2-3**. Effects of practice variability on QRISK3 linear predictor (random slope) (60% of overall CPRD practices)

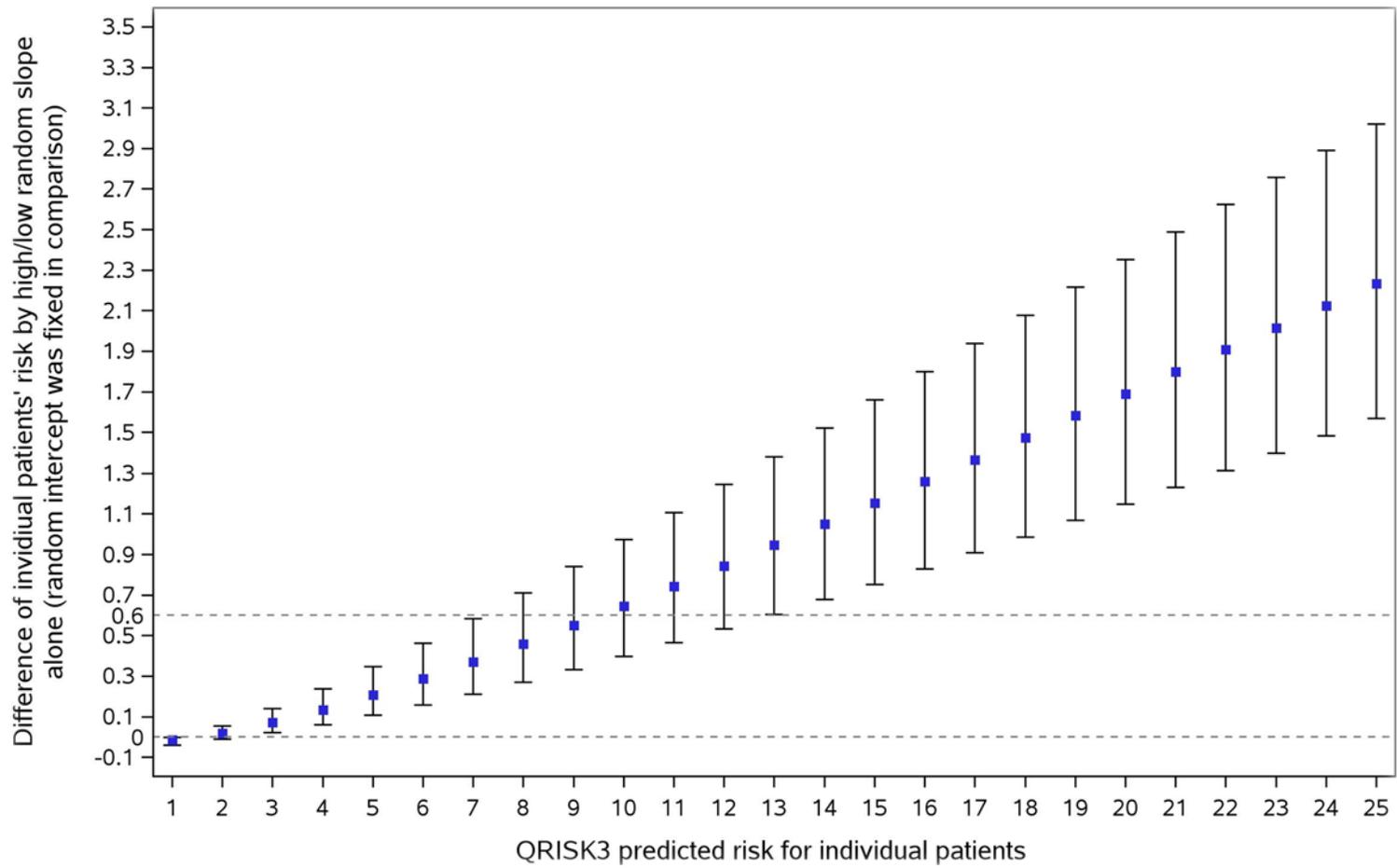

**eFigure 3. Difference of individual patients' prediction between practice with 2.5% random slope and 97.5% slope and a random selected fixed random intercept**